\documentclass[prl,amsmath,twocolumn,superscriptaddress,showpacs]{revtex4}
\usepackage{bm}
\usepackage{amssymb}
\usepackage{graphicx}

\newcommand{\be}{\begin{equation}}
\newcommand{\ee}{\end{equation}}
\newcommand{\bea}{\begin{eqnarray}}
\newcommand{\eea}{\end{eqnarray}}
\newcommand{\bsube}{\begin{subequations}}
\newcommand{\esube}{\end{subequations}}

\begin{document}

\title{Magnetic field switching in parallel quantum dots}

\author{Feng Li}
\affiliation{State Key Laboratory for Superlattices and
Microstructures, Institute of Semiconductors, Chinese Academy of
Sciences, P.O.~Box 912, Beijing 100083, China}

\author{Xin-Qi Li}
\affiliation{State Key Laboratory for Superlattices and
Microstructures, Institute of Semiconductors, Chinese Academy of
Sciences, P.O.~Box 912, Beijing 100083, China}

\author{Wei-Min Zhang}
\affiliation{National Center for Theoretical Science, Tainan, Taiwan
70101, ROC}

\affiliation{Department of Physics and Center for Quantum
Information Science, National Cheng Kung University, Tainan, Taiwan
70101, ROC}

\author{S.A. Gurvitz}
 \affiliation{Department of Particle Physics,
 Weizmann Institute of Science, Rehovot 76100, Israel
 }

\date{\today}
\pacs{73.23.-b, 73.23.Hk, 05.60.Gg}

\begin{abstract}
We show that the Coulomb blockade in parallel dots pierced by
magnetic flux $\Phi$ completely blocks the resonant current for any
value of $\Phi$ except for integer multiples of the flux quantum
$\Phi_0$. This non-analytic (switching) dependence of the current on
$\Phi$ arises only when the dot states that carry the current are of
the same energy. The time needed to reach the steady state, however,
diverges when $\Phi\to n\Phi_0$.
\end{abstract}

\maketitle
\begin{figure}
\includegraphics[width=8cm]{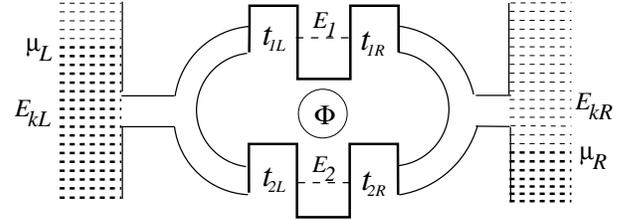}
\caption{Resonant tunneling through two parallel dots pierced by the
magnetic flux $\Phi$.} \label{fig1}
\end{figure}

 The system of two quantum dots coupled in parallel to two
reservoirs has attracted a great deal of attention as a realization
of a mesoscopic Aharonov--Bohm interferometer \cite{holl,chen,sigr}.
Indeed such a system pierced by an external magnetic field
(Fig.~\ref{fig1}) is an interference device whose transmission can
be tuned by varying the magnetic field. In the absence of the
interdot electron--electron interaction, the interference effects in
the resonant current through this system are quite transparent. This
is not the case, however, for interacting electrons
\cite{gefen,neder}.

Consider for instance a strong interdot electron--electron
repulsion---a Coulomb blockade. While the two dots may be occupied
simultaneously in the noninteracting model, the Coulomb blockade
prevents this. At first sight one might not expect that this
repulsion could dramatically modify the resonant current's
dependence on the magnetic field. We find, however, that the
resonant current is completely blocked for any value of the magnetic
flux except for integer multiples of the flux quantum $\Phi_0=h/e$.
This striking effect has so far been overlooked in the literature,
despite the current-switch in tunnel-coupled quantum dots, induced
by coherent radiation of two microwaves, was proposed earlier \cite
{brandes}.

In this Letter we show how such a switching effect can be seen in
the exact analytical solution for certain symmetric choices of the
quantum dots' parameters. This solution can be found in the infinite
bias limit, which we justify by comparing the finite- and
infinite-bias cases for non-interacting electrons. We then present
an explanation of the switching effect as a self-trapping phenomenon
in non-linear systems. This explanation allows us to formulate
general conditions for this phenomenon and clearly displays its
generic nature.

Consider a double dot (DD) connected in parallel to two reservoirs,
as shown in Fig.~\ref{fig1}. For simplicity we consider spinless
electrons. We also assume that each of the dots contains only one
level, $E_{1}$ and $E_{2}$ respectively. In the presence of a
magnetic field, the system can be described by the following
tunneling Hamiltonian,
\begin{align}
H=H_0+H_T+\sum_{\mu =1,2} E_\mu d_\mu^\dagger d_\mu +Ud_1^\dagger
d_1 d_2^\dagger d_2\, . \label{a1}
\end{align} Here $H_0=\sum_k [E_{kL}a_{kL}^\dagger a_{kL}
+E_{kR}a_{kR}^\dagger a_{kR}]$ and $H_T$ describes the reservoirs
and their coupling to the dots,
\begin{align}
H_T=\sum_{\mu,k}\Big (t_{\mu L}d_\mu^\dagger a_{kL} +t_{\mu
R}a_{kR}^\dagger d_\mu\Big )+H.c.\, , \label{a2}
\end{align}
where $\mu=1,2$ and $a_{kL(R)}^\dagger$ are the creation operators
for the electrons in the reservoirs while $d_{1,2}^\dagger$ are the
creation operators for the DD. The last term in Eq.~(\ref{a1})
describes the interdot repulsion. We assume that there is no direct
transmission between the dots and that the couplings of the dots to
the leads, $t_{\mu L(R)}$, are independent of energy. In the absence
of a magnetic field all couplings are real. In the presence of a
magnetic flux $\Phi$, however, the tunneling amplitudes between the
dots and the reservoirs are in general complex. We write $t_{\mu
L(R)}={\bar t}_{\mu L(R)}e^{i\phi_{\mu L(R)}}$, where $\bar t_{\mu
L(R)}$ is the coupling without the magnetic field. The phases around
the closed circle are constrained to satisfy
$\phi_{1L}+\phi_{1R}-\phi_{2L}-\phi_{2R}=\phi$, where $\phi\equiv
2\pi\Phi/\Phi_0$.

Let the initial state of the system correspond to filling the left
and right reservoirs at zero temperature with electrons up to the
Fermi energies $\mu_L$ and $\mu_R$, respectively (Fig.~\ref{fig1}).
Consider first noninteracting electrons, $U=0$. In this case the
problem can be solved exactly for any values of the bias voltage,
$\mu_L-\mu_R$, and of the couplings to the leads \cite{g1}. Indeed,
the total wave function for the noninteracting electrons,
$|\Psi(t)\rangle=\exp (-iHt)|\Psi(0)\rangle$, can be written at all
times as a product of single-electron wave functions, $|\Psi
(t)\rangle=\prod_{k'} |\psi_{k'}(t)\rangle$.  Here
$|\psi_{k'}(t)\rangle$ describes a single electron initially
occupying one of the levels $E_{k'L} \le \mu_L$, or $E_{k'R} \le
\mu_R$ in the left or right lead. It can be written as
\begin{align}
|\psi_{k'}(t)\rangle=\Big[\sum_{k,\alpha=L,R}
b_{k'k}^{\alpha}(t)a_{k\alpha}^\dagger
 +&\sum_{\mu=1,2}
b_{k'\mu} (t)d_\mu^\dagger \Big]|0\rangle\, , \label{a3}
\end{align}
where $b_{k'k}^{L(R)}(t)$ and $b_{k'\mu}(t)$ are the probability
amplitudes for finding the electron at the level $E_{kL(R)}$ in the
left (right) reservoir, or at the level $E_\mu$ inside the DD system.
The total probability of finding the
electron in the right lead is therefore
$P_{k'}(t)=\sum_k|b_{k'k}^R(t)|^2$

Consider $\mu_L>\mu_R$. Then the total average charge $Q(t)$
accumulated in the right lead by time $t$ is a sum of $P_{k'}(t)$
over all electrons with initial energy within the potential bias,
$\mu_R<E_{k'L}<\mu_L$. The average current is $I(t)=\dot{Q}(t)$. (We
adopt units where the electron charge $e=1$). Let us take the
continuum limit $\sum_k\to\int\rho\, dE_k$, where $\rho$ is the
density of state in the corresponding lead. Then the current can be
written as
\begin{align}
I(t)=\int_{\mu_R}^{\mu_L}\rho_LI_{k'}(t)dE_{k'L}\, ,
\label{a4}\end{align} where
$I_{k'}(t)=\partial_t\int_{-\infty}^\infty |b^R_{k'k}(t)|^2dE_{kR}$
is a single electron current.

Substituting Eq.~(\ref{a3}) into the Schr\"odinger equation
$i|\dot{\psi}_{k'}(t)\rangle =H|\psi_{k'}(t)\rangle$ we obtain the
following equations for the amplitudes $b(t)$,
\begin{subequations}
\label{SE-1}
\begin{align}
i\dot{b}^L_{k'k} &=E_{kL} b^L_{k'k} + \sum_{\mu=1,2}
                    t^*_{\mu L}b_{k' \mu}
   \label{SE-1a}\\
i\dot{b}_{k'\mu} &=E_{\mu} b_{k'\mu} + \sum_k\left(t_{\mu L}
 b^L_{k'k}+t^*_{\mu R}  b^R_{k'k}
\right)\label{SE-1b}\\
i\dot{b}^R_{k'k} &=E_{kR} b^R_{k'k} + \sum_{\mu=1,2}
                    t_{\mu R} b_{k'\mu} \, .\label{SE-1c}
\end{align}
\end{subequations}
Replacing sum over the lead states by integrals, these equations can
be solved analytically. Taking for simplicity $\bar t_{1L(R)}=\bar
t_{2L(R)}\equiv \bar t_{L(R)}$ we obtain for a single electron
stationary current ($t\to\infty$) the following result:
\begin{align} I_{k'}
  = 4\Gamma_R |t_L|^2 \left[~|f_1|^2+|f_2|^2
      +2 {\rm Re} (e^{-i\phi}f_1f_2) \right]/
   D^2\label{a5}
\end{align} where $\Gamma_{L(R)}=2\pi \rho_{L(R)} |t_{L(R)}|^2$ and
\begin{align}  &f_{1,2}
    = (E_{k'L}-E_{2,1})\mp\frac{i}{2}(\Gamma-\Gamma_{\phi})\nonumber\\
&D=\left( 2E_{k'L}-E_1-E_2+ i\Gamma \right)^2-
\epsilon^2+|\Gamma_{\phi}|^2\, .\nonumber
\end{align}
Here $\varepsilon =E_1-E_2$, $\Gamma =\Gamma_L+\Gamma_R$, and
$\Gamma_{\phi} = \Gamma_L + \Gamma_R e^{i\phi}$.

\begin{figure}
\includegraphics[width=8cm]{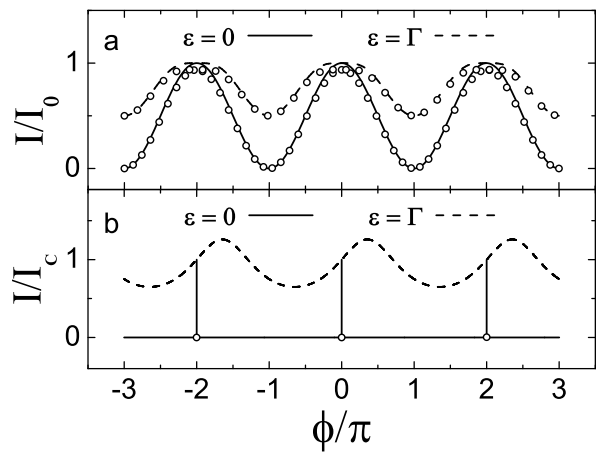}
\caption{(a) Stationary current versus magnetic flux for the
noninteracting DD with aligned and misaligned levels, where
$\Gamma_L=\Gamma_R=\Gamma/2$. Open circles show the result of
numerical evaluation of Eqs.~(\ref{a4}), (\ref{a5}) for a finite
voltage $\mu_{L,R}=E_{1,2}\pm 5\,\Gamma$; the curves show
Eq.(\ref{a6}), corresponding to infinite bias. (b) Stationary
current for the same system in the Coulomb blockade regime,
Eq~(\ref{a10}).} \label{fig2}
\end{figure}

In the case of large bias, $|\mu_{L,R}-E_{1,2}|\gg\Gamma$, the
integration limits in Eq.~(\ref{a4}) can be extended to infinity. As
a result, we find for the total current $I\equiv I(\phi)$,
\begin{align} I(\phi ) =
I_0{\varepsilon^2+\Gamma_L\Gamma_R\sin^2 \phi
\over\varepsilon^2+4\Gamma_L\Gamma_R\sin^2{\displaystyle\phi\over\displaystyle
2}\, }\, , \label{a6}
\end{align}
where $I_0=2\Gamma_L\Gamma_R/\Gamma$ is the resonant current for
non-interacting electrons in the absence of the magnetic filed. The
$\phi$-dependence in Eq.(\ref{a6}) is an example of the
Aharonov--Bohm effect; we illustrate this for finite and infinite
bias in Fig.~\ref{fig2}a. We find the infinite bias limit
Eq.~(\ref{a6}) is a very good approximation for a finite bias
whenever the level is well inside the bias window,
$|\mu_{L,R}-E_{1,2}|\gg\Gamma$.

The entire treatment can in fact be simplified in the large bias
limit by transforming Eqs.~(\ref{SE-1}) into Bloch-type master
equations for the reduced density matrix of the DD system,
$\sigma_{jj'}(t)$. Here $j(j')=\{0,1,2,3\}$ label the DD states in
order: the empty DD, the first dot occupied, the second dot
occupied, and both dots occupied. This density matrix is related to
the amplitudes $b_{k'\mu}(t)$ via~\cite{gmb}
\begin{align}
\sigma_{\mu\mu'}(t)+\delta_{\mu\mu'}\sigma_{33}(t)=\int_{-\infty}^\infty
b_{k'\mu}(t)b^*_{k'\mu'}(t)\rho_LdE_{k'L} \label{aa1} \end{align}
 for $\mu
=1,2$ and $\sigma_{00}=1-\sigma_{11}-\sigma_{22}-\sigma_{33}$. Then
multiplying Eqs.~(\ref{SE-1}) by $b_{k'\mu}^*(t)$ and integrating
over $E_{k'L}$ gives the master equations
\begin{subequations}
\label{a7}
\begin{align}
&\dot{\sigma}_{00}= -2 \Gamma _L \sigma_{00}+\Gamma _R (\sigma
_{11}+\sigma_{22}+\bar\sigma_{12}+
\bar\sigma_{21})\label{a7a}\\
&\dot{\sigma}_{11}= \Gamma _L
\sigma_{00}-\Gamma\sigma_{11}+\Gamma_R\sigma_{33}+\delta\Gamma^*
\bar\sigma_{12} +\delta\Gamma\bar\sigma_{21}
\label{a7b}\\
&\dot{\sigma}_{22}= \Gamma _L
\sigma_{00}-\Gamma\sigma_{22}+\Gamma_R\sigma_{33}+\delta\Gamma^*
\bar\sigma_{12} +\delta\Gamma\bar\sigma_{21}
\label{a7c}\\
&\dot{\sigma}_{33}=\Gamma_L(\sigma_{11}+\sigma _{22}-e^{-i\phi}
\bar\sigma_{12}-e^{i\phi}\bar\sigma_{21})-2\Gamma
_R\sigma_{33}\label{a7d}\\
&\dot{\bar\sigma}_{12}=e^{i\phi} \Gamma _L \sigma_{00}+\delta\Gamma(
\sigma_{11}+\sigma_{22})-\Gamma _R \sigma_{33}\nonumber\\
&~~~~~~~~~~~~~~~~~~~~~~~~~~~~~~~~~~~~~~~~~~~~~~-(i\varepsilon
+\Gamma )\bar\sigma_{12}\label{a7e}
\end{align}
\end{subequations}
where
$\bar\sigma_{12}(t)=\exp[i(\phi_{1R}-\phi_{2R})]\sigma_{12}(t)$ and
$\delta\Gamma =(e^{i\phi}\Gamma_L-\Gamma_R)/\, 2$. The total current
is given by \cite{gmb}
\begin{align}
I(t)=\Gamma_R[\sigma_{11}(t)+\sigma_{22}(t)+2\sigma_{33}(t)+2{\rm
{Re}}\,\bar\sigma_{12}(t)]. \label{a8}
\end{align}
Solving Eqs.~(\ref{a7}) in the steady-state limit ($\dot{\sigma}\to
0$ for $t\to\infty$) we easily reproduce Eq.~(\ref{a6}).

In contrast with the single-electron approach, the master equations
(\ref{a7}) can be applied only for large bias. On the other hand,
the master equations can accommodate the Coulomb blockade case
($E_{1,2}+U\gg \mu_L$) in the most simple way. Indeed, the Coulomb
blockade merely leads one to exclude the states corresponding to a
simultaneous occupation of the two dots from Eqs.~(\ref{a7}). As a
result the master equations become
\begin{subequations}
\label{a9}
\begin{align}
&\dot{\sigma}_{00}= -2 \Gamma _L \sigma_{00}+\Gamma _R (\sigma
_{11}+\sigma_{22}+\bar\sigma_{12}+
\bar\sigma_{21})\label{a9a}\\
&\dot{\sigma}_{11}= \Gamma _L
\sigma_{00}-\Gamma_R\sigma_{11}-\Gamma_R( \bar\sigma_{12}
+\bar\sigma_{21})/2
\label{a9b}\\
&\dot{\sigma}_{22}= \Gamma _L
\sigma_{00}-\Gamma_R\sigma_{22}-\Gamma_R( \bar\sigma_{12}
+\bar\sigma_{21})/2
\label{a9c}\\
&\dot{\bar\sigma}_{12}=e^{i\phi} \Gamma _L
\sigma_{00}-{\Gamma_R\over2} (\sigma_{11}+\sigma_{22})-(i\varepsilon
+\Gamma_R )\bar\sigma_{12} \label{a9d}
\end{align}
\end{subequations}
where the total current is given by Eq.~(\ref{a8}) with
$\sigma_{33}=0$. Equations~(\ref{a9}) can in fact be rigorously
derived by employing the nonequilibrium Green's function in the weak
coupling limit \cite{gefen,dong}, or directly from the many-body
Schr\"odinger equation in the large bias limit \cite{gp} by assuming
that $\max(\Gamma,T)/|\mu_{L,R}-E_{1,2}|\ll 1$, where $T$ is the
temperature of the reservoirs~\cite{li,li1}. Under this assumption
they are valid to all orders in the tunneling couplings. (In fact,
the master equations approach has been already shown to be very
useful for description of electron transport in coupled dots
\cite{brandes,sauret,dong1}).

Solving Eqs.~(\ref{a9}) in the steady-state limit we obtain for the
total current
\begin{align}
I(\phi )=I_C{\varepsilon^2\over\varepsilon^2+I_C\left(2\Gamma_R
\sin^2{\displaystyle \phi\over\displaystyle 2} -\varepsilon\sin \phi
\right)}\, , \label{a10}\end{align} where
$I_C=2\Gamma_L\Gamma_R/(2\Gamma_L+\Gamma_R)$ is the total current
(with the Coulomb blockade) in the absence of the magnetic field.
Comparing Eq.~(\ref{a10}) with Eq.~(\ref{a6}) for $\varepsilon\not
=0$ we find that both currents display the Aharonov--Bohm
oscillations, Fig.~\ref{fig2}. Nevertheless, Eq.~(\ref{a10}) shows
an asymmetric behavior with respect to the magnetic flux, $\phi$
(the dashed curve in Fig.~\ref{fig2}b). It looks contradictory to
the Onsager relation that locks the current peaks at $\phi=2\pi n$
in any \emph{two-terminal linear} transport \cite{Buttiker}. Yet,
this relation is not applicable in our case, corresponding to
interacting system {\em under finite bias voltage} \cite{gefen}.

More strikingly is the current behavior in the interacting case for
$\varepsilon \to 0$. While the resonant current for the
noninteracting electrons keeps oscillating with the magnetic field,
Fig.~\ref{fig2}a, it becomes non-analytic in $\phi$ in the case of
Coulomb blockade. Indeed, one finds from Eq.~(\ref{a10}) that
$I=I_C$ for $\phi =2\pi n$, where $n=\Phi /\Phi_0$ is an integer,
but $I=0$ for any other value of $\Phi$ (Fig.~\ref{fig2}b).

Such an unexpected ``switching'' behavior of the electron current in
the magnetic field can be explained in the following way. Let us
disentangle the Coulomb blockade and quantum interference effect,
which interplay in a non-trivial way in the electron current. This
can be done by defining new basis states of the DD,
$d^\dagger_\mu|0\rangle\to\widetilde d^\dagger_\mu |0\rangle$,
chosen so that they do not interfere in the electron current. For
instance, if the state $\widetilde d^\dagger_2 |0\rangle$ is not
coupled to the right reservoir, {\em i.e.}, $t_{2R}\to \tilde
t_{2R}=0$, then the current would flow only through the state
$\widetilde d_{1}^\dagger|0\rangle$. Obviously, no interference
between these two states would appear in the total current.

Such a basis can be found for $\varepsilon =E_1-E_2=0$ when the DD
Hamiltonian, $E_1d_1^\dagger d_1+E_2d_2^\dagger d_2$, is invariant
under SU(2) transformations. Then the unitary transformation
\begin{align}
\left (\begin{array}{c}\widetilde {d}_{1}\\
\widetilde {d}_{2}\end{array}\right)={1\over {\cal
N}}\left(\begin{array}{cc}t_{1R}
& t_{2R}\\
-t^*_{2R}&t^*_{1R}\end{array}\right)\left (\begin{array}{c}d_{1}\\
d_{2}\end{array}\right), \label{a11}
\end{align}
with ${\cal N}=(\bar t_{1R}^2+\bar t_{2R}^2)^{1/2}$, results in
$\widetilde t_{2R}=0$.

Consider now the coupling of the state  $\widetilde d^\dagger_2
|0\rangle$ to the left lead. One obtains from Eq.~(\ref{a2}) for
$\widetilde t_{2L}\equiv \widetilde t_{2L}(\phi)$
\begin{align}
\widetilde t_{2L}(\phi)=-e^{i(\phi_{2L}-\phi_{1R})}(\bar t_{1L}\bar
t_{2R}\, e^{i\phi}-\bar t_{2L}\bar t_{1R})/{\cal N} .\label{a12}
\end{align}
It follows from this expression that $\widetilde t_{2L}=0$ for
$\phi=2n\pi$ provided that $\bar t_{1L}/\bar t_{2L}=\bar t_{1R}/\bar
t_{2R}$, or for $\phi=(2n+1)\pi$ if $\bar t_{1L}/\bar t_{2L}=-\bar
t_{1R}/\bar t_{2R}$ \footnote{The latter takes place for
non-identical dots where $E_1=E_2$ and one of these states is an
excited state, see G. Hackenbroich, Phys. Rep. 343, 463 (2001).}.
Then the state $\widetilde d_2^\dagger |0\rangle$ decouples from
both leads. This would not affect the resonant current for
noninteracting electrons ($U=0$), since the state $\widetilde
d_2^\dagger |0\rangle$ is already decoupled from the right lead.

In the case of Coulomb blockade, however, the coupling to the left
lead becomes the point of crucial importance. Indeed, any discrete
state coupled to an infinite reservoir is going to be totally
occupied, no matter how weak the coupling \cite{g1}. Then the state
$\widetilde d_1^\dagger |0\rangle$, carrying the current, will be
blocked by the Coulomb interdot repulsion. As a result, the total
current {\em vanishes}. However, if the state $\widetilde
d_2^\dagger |0\rangle$ is decoupled from {\em both\/} leads, it
remains unoccupied, so that the current can flow through the state
$\widetilde d_1^\dagger |0\rangle$. As shown above, this takes place
precisely for $\bar t_{1L}/\bar t_{2L}=\pm\,\bar t_{1R}/\bar
t_{2R}$. If this condition is not fulfilled, the current is always
zero, even for $\phi =2\pi n$.

It is clear from our explanation that the switching phenomenon
disappears for $\varepsilon\not =0$, as can be seen from
Eq.~(\ref{a10}). Indeed, in this case the DD Hamiltonian $\sum_\mu
E_\mu d^\dagger_\mu d_\mu$ is not invariant under SU(2)
transformations. Therefore any unitary transformation that
eliminates one link between the DD and the reservoirs, like
Eq.~(\ref{a11}), would generate direct coupling ($\tilde
t_{12}=\varepsilon\, t_{1R}t_{2R}/[|t_{1R}|^2+|t_{2R}|^2]$) between
the states $\widetilde d_{1,2}^\dagger |0\rangle$\footnote{No such
coupling is produced by the interdot repulsion term in
Eq.~(\ref{a1}), since it is always invariant under the
transformation (\ref{a11}): $Ud_1^\dagger d_1d_2^\dagger d_2\to
U\widetilde d_1^\dagger \widetilde d_1\widetilde d_2^\dagger
\widetilde d_2$.}. As a result, the current is not interrupted by
the Coulomb blockade, so that its behavior for $\varepsilon\not =0$
would be similar to that in the noninteracting case,
Fig.~\ref{fig2}.

\begin{figure}
\includegraphics[width=7cm]{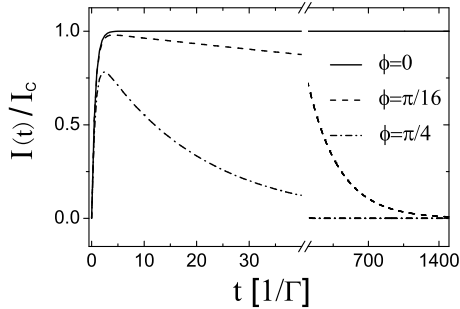}
\caption{Transient behavior of the electric current  for different
values of the magnetic flux, obtained from Eqs.~(\ref{a9}).}
\label{fig3}
\end{figure}

The switching effect of the magnetic field on the electric current
takes place only in the stationary regime, $t\to\infty$, where one
of the states $\widetilde d_{1,2}^\dagger |0\rangle$ is fully
occupied. However, it takes a certain transition time, $t_{\rm tr}$,
during which the total current is not zero. One estimates from
Eq.~(\ref{a9}) that $t_{\rm tr}\sim 1/\widetilde{\Gamma}_{2L}$,
where $\widetilde{\Gamma}_{2L}=2\pi\rho_{L}|\widetilde{t}_{2L}|^2$.
Using Eq.~(\ref{a12}) we find that $t_{\rm tr}\sim
\Gamma^{-1}(\Phi_0/\Phi )^2$ for $\Phi\ll\Phi_0$. Hence $t_{\rm tr}$
becomes much longer than the usual relaxation time $\Gamma^{-1}$ for
a very small magnetic field (or in general when $\Phi\to n\Phi_0$).
This is illustrated in Fig.~\ref{fig3}, which shows the transient
current $I(t)$ as a function of $t$ for different values of $\phi$.
One finds from this figure that the current always increases for
small $t$. However, it eventually disappears for $t\gg t_{\rm
{tr}}$. This explains the non-analyticity of $I(\phi )$ at $\phi
=2\pi n$. Indeed, at any finite time $t$ there is no discontinuity
at $\phi=2\pi n$. It appears only in the limit of $t\to\infty$
because $t_{\rm tr}\to\infty$ for $\Phi\to 0$.

As we demonstrated above, the switching effect becomes very
transparent in a particular basis of the DD states. Still, it is
very surprising how such a basis emerges dynamically? Indeed, an
electron from the left lead can enter the DD system in any of SU(2)
equivalent superpositions of its states. Therefore there exists a
probability for each electron to enter the DD in the superposition
that eliminates one of the links with the right lead. When it
happens, the electron would be trapped in this state. Even if the
probability of this event for one electron is very small, the total
number of electrons passing the DD goes to infinity for
$t\to\infty$. Therefore the trapping event is always realized for
large enough time. In the presence of Coulomb blockade this would
lead to the switching effect, as explained above. (A similar effect
of the Coulomb blockade, leading to divergency in the shot-noise
power has been discussed in different publications
\cite{dong1,burkard,urban,li2}. However, this phenomenon is not
related to vanishing of current, but rather to electron bunching
leading to a system's bistability \cite{urban,li2}.)

Our interpretation of the switching effect allows us to determine
the necessary conditions for its realization in real experiment.
First, we need the total occupation of one of the states $\widetilde
d_{1,2}^\dagger |0\rangle$, decoupled with the right lead. This
takes place only if the energy level is far from the corresponding
Fermi energy, $|\mu_L-E_1|\gg\Gamma$. For instance, if $E_1=\mu_L$,
the occupation probability reaches only $1/2$ at $t\to\infty$. As
the bias increases, however, other levels of the dots can enter into
the bias window. In this case the transport would proceed through
several levels. If the dots are identical, the same ``rotation''
(\ref{a11}), applied to each pair of levels with the same energy,
would result in a simultaneous decoupling of the corresponding
``rotated'' states from the right lead. Then the switching effect is
expected to take place in this case as well, even if the inter-level
spacing is very small.

In real system, the surrounding environment will cause dephasing
between the two dots due to fluctuations of the dots parameters,
like energy levels, tunneling couplings, etc. Since a particular
origin of these fluctuation is irrelevant for evaluation of the
corresponding dephasing rate \cite{gur4}, one can model the
environment by isolated fluctuators interacting with the system and
vibrating its energy levels\cite{li2}. Using such a model for the
fluctuating environment one finds for the stationary current (for
$\varepsilon =0$ and $\Gamma_L=\Gamma_R=\Gamma /2$) \cite{li2},
\begin{align}
I={I_c\over 1+I_c\tau_d\sin^2(\phi/2)}\, , \label{in1}
\end{align}
where $I_c=\Gamma/3$ and $\tau_d$ the dephasing time. Therefore in
order to observe the switching effect one requires $I_c\tau_d \gg
1/\sin^2(\phi /2)$. For small $\phi$ it is equivalent to $\tau_d\gg
t_{\rm tr}$.

In general, for a weak coupling to the environment, the dephasing
rate $\gamma_d=1/\tau_d$ can be evaluated as $\gamma_d\sim
(\delta\epsilon)^2S(0)$, where $\delta\epsilon$ is an average
fluctuation of dots levels and $S(\omega )$ is the corresponding
spectral density \cite{gur4,legg}. In the case of the thermal
environment $S(\omega )\propto T$, where $T$ is temperature
\cite{legg}. Therefore by decreasing the temperature one can make
the corresponding decoherence rate arbitrary small in order to reach
the switching effect even for small values of the magnetic flux.

As we explained above the experimental realization of the switching
effect would require fulfillment of different conditions which
should be met. The most essential of them are large bias and
inter-dot Coulomb repulsion, which should exceed the bias.

With respect to the bias, it would be hard to realize the large bias
voltage (on scale of $\Gamma$) with one level inside the bias
window. However, in the case of two identical dots, the switching
effect is expected even if many levels are inside the bias.
Therefore the large bias condition should not create essential
experimental problem.

A realization of the large inter-dot repulsion condition would
represent a more complicated problem. Indeed, it implies that two
dots are very close. This condition is still not met in present
experiments \cite{holl} . However, it can be met by decreasing the
dots size.  An another way to achieve proximity of two parallel dots
is to use different materials. For instance the quantum dots in
graphene system can be a very promising set-up for an investigation
of the switching effect \cite{ens}.

\begin{acknowledgments} This work was supported by the National
Natural Science Foundation of China under grants No.\ 60425412 and
No.\ 90503013, the Major State Basic Research Project under grant
No.2006CB921201. X.Q.L. acknowledges the Einstein center for
partially supporting his visit to the Weizmann Institute of Science.
Two of us (X.Q.L.) and (S.A.G.) are grateful to NCTS, Tainan, Taiwan
for kind hospitality. We also thank B. Svetitsky for important
suggestions to this paper.
\end{acknowledgments}

\end{document}